\documentclass[prd,showpacs,twocolumn,amsfonts]{revtex4}
\usepackage{epsfig}
\usepackage{graphics}

\newcommand{\Eqs}[2]{Eqs.~(\ref{#1}) and (\ref{#2})}
\newcommand{\Eq}[1]{Eq.~(\ref{#1})}

\newcommand{\beq}{\begin{equation}}
\newcommand{\eeq}{\end{equation}}
\newcommand{\bea}{\begin{eqnarray}}
\newcommand{\eea}{\end{eqnarray}}
\def\half{{\scriptstyle{\frac{1}{2}}}}
\def\fourth{{\scriptstyle{\frac{1}{4}}}}

\begin{document}

\title{Varying  fine structure ``constant'' and charged black holes}

\author{Jacob D. Bekenstein}\affiliation{Hebrew University of Jerusalem, Jerusalem 91904, Israel}
\author{Marcelo Schiffer}\affiliation{Ariel University Center of Samaria, Ariel 44837, Israel}

\date{\today}

\begin{abstract}
Speculation that the fine-structure constant $\alpha$ varies in spacetime has a long history.  We derive, in 4-D general relativity and in isotropic coordinates, the solution for a charged spherical black hole according to the framework for dynamical $\alpha$ (Bekenstein 1982).   This solution coincides with a previously known one-parameter extension of the dilatonic black hole family.   Among the notable properties of varying-$\alpha$ charged black holes are adherence to a ``no hair'' principle,  the absence of the inner (Cauchy) horizon of the Reissner-Nordstr\" om black holes, the nonexistence of precisely extremal black holes, and the appearance of naked singularities in an analytic extension of the relevant metric.  The exteriors of almost extremal electrically (magnetically) charged black holes have simple structures which makes their influence on applied magnetic (electric) fields transparent.  We re-derive the thermodynamic functions of the modified black holes; the otherwise difficult calculation of the electric potential is done by a shortcut.  We confirm that variability of  $\alpha$ in the wake of expansion of the universe does not threaten the generalized second law
\end{abstract}


\maketitle 

\section{\label{sec:intro}Introduction}

The issue of temporal variability of the fine-structure
constant $\alpha$, first considered theoretically by
Jordan~\cite{jordan,uzan}, Teller~\cite{teller}, Gamow~\cite{gamow}, Dicke~\cite{dicke} and Stanyukovich~\cite{stan}, has assumed added urgency in view of controversial
claims of cosmological variability in the
fine-structure multiplets in quasar spectra~\cite{webb}  (for reviews of the issue see Refs.~\cite{uzan,wetterich}).
Jordan and  Dicke emphasized
that in a covariant theory of variable $\alpha$, temporal  variability
must be accompanied by spatial variability.   In the model
theory exhibited by Dicke---really the first theory with just $\alpha$ varying---this spatial
variation seems to clash with the equivalence principle, an issue to which we shall return. 

Dicke's model for $\alpha$-variability
determines the full $\alpha$
field rather than just  its relative variation in spacetime---the theory's action is not invariant under rescaling of the relevant field by a constant factor.    However,  a theory of variability of a coupling
constant, like the fundamental charge, should not determine
its overall scale, since the later has to do with the system of global units employed.  Put another way, a theory of a varying electric charge can be rewritten as a theory with a fixed charge but with
varying electric permittivity of the vacuum~\cite{dicke,bek}.  Again, such theory should not fix the overall scale of the permittivity, which depends on the units adopted for the electric field and displacement.  

A framework that determines the variation of $\alpha$, but not its overall scale,  was proposed by one of us some years ago~\cite{bek,bek2}, and explored exhaustively by Magueijo, Barrow and Sandvik among others~\cite{MBS}.  It can be written as a theory of varying charge~\cite{bek}, or as one of varying vacuum permittivity~\cite{bek2}.  This framework assumes that for constant $\alpha$ electromagnetism reduces to
Maxwell's with  minimal coupling to charged matter, that
$\alpha$ dynamics comes from an action which, like the
Maxwellian action, is coordinate and gauge invariant,
that the theory is classically causal and respects
time reversal invariance, that any length scale in the
theory is no smaller than Planck's length
$L_{\rm P}=(\hbar G/c^3)^{1/2}\approx 1.616\times 10^{-33}$ cm,
and that gravitation is governed by the Einstein-Hilbert
action.  

We shall describe the framework in the language of a varying vacuum permittivity~\cite{MBS,bek2}.   This last is supposed to be represented by 
$e^{-2\psi}$ where $\psi$ is a real scalar
field.  The electromagnetic field tensor $F_{\mu\nu}$
is supposed to be derivable from a 4-potential.  Thus
\beq
{}^*F^{\mu\nu}{}_{;\nu}=0,
\label{dual}
\eeq
where the dual is defined in terms of the Levi-Civitta tensor by ${}^*\! F^{\mu\nu} =
\half
\epsilon^{\mu\nu\alpha\beta} F_{\alpha\beta}$.  The dynamics
of $\psi$ and $F^{\mu\nu}$ are governed by the combined
action
\beq
S=-{1\over 16\pi}\int \Big(e^{-2\psi}
F_{\mu\nu}F^{\mu\nu}+{2\over
\kappa^{2}}\psi_{,\mu}\psi_,{}^{\mu}\Big)(-g)^ {1/2}d^4x,
\label{action}
\eeq
which goes over to Maxwell's in the limit $\psi\rightarrow
\textrm{const.}$, as required.  Here $\kappa\equiv l(4\pi\hbar
c)^{-1/2}$ with $l$ a length parameter of the theory.    Note that a shift $\psi\to\psi+$ const. merely ``renormalizes''  the electromagnetic part of $S$; it can thus be construed as consequence of a change of the electromagnetic  units.  This accords with the characterization of the framework as one that does not  determine the overall scale of the permittivity. To the stated action one must add the Einstein-Hilbert one for gravity (which we do not display) and the standard one for the matter.  Being interested in bare black holes we drop the latter.

 This is the place to remark on the close similarity (but not identity) between the framework and the so called dilatonic sector of the low energy limit of string theory (dilaton theory)~\cite{gibb,GM,ghs,CKMO}.  In dilaton theory the equivalent of the framework's length scale $l$ must be $L_{\rm P}$, but in the framework $l$ is a free parameter to be determined experimentally.  The framework  was thought to predict~\cite{bek}  that with $l \geq  L_{\rm P}$ departures from the weak equivalence principle occur which would have been ruled out by the classic experiment of Dicke et al~\cite{RKD}.  Olive and Pospelov~\cite{OP} investigated a modification of the framework designed to eliminate this problem ($\psi$ field couples more strongly to dark matter than to baryonic).  However, more detailed analysis~\cite{LSV,bek2} has suggested that equivalence principle violations in the original framework may actually be suppressed (see Ref.~\onlinecite{Damour} for a differing opinion).  

The fact that $\alpha$ variability goes hand in hand with a modification of Maxwellian electrodynamics means that the charged black holes in the framework must be distinct from the Reissner-Nordstr\"om (RN) and Kerr-Newman families of solutions of the Einstein-Maxwell theory.  A family of spherically symmetric charged black holes of the dilaton theory which supplant the RN black holes has long been known~\cite{gibb,GM,ghs,CKMO} (these \emph{dilatonic black holes} have recently been extended to take into account phantom matter sources~\cite{CFR}). But it was not clear that the dilatonic black holes are the unique spherical black hole family in face of $\alpha$ variability.  Neither were the coordinates in which the metric for a dilatonic black hole is usually couched particularly conducive to visualization of its geometry.  We have thus here derived \emph{ab initio}, and in standard isotropic coordinates, the \emph{unique} geometry and electric field of a spherical static charged black hole within the variable $\alpha$ framework.  We show them to be those of a dilatonic black hole (by transforming to the coordinates used by Garfinkle, Horowitz and Strominger (GHS)~\cite{ghs}).  We elucidate the geometric, electrodynamic and thermodynamic properties of the modified black holes, including some that were previously unknown.  

In Sec.~\ref{sec:equations} we collect the equations to be solved for a static spherical system with electromagnetism as the only matter source.  In Sec.~\ref{sec:BH} we delineate the expected features expected of a black hole solution, and in light of them we solve the aforementioned equations to obtain the generic black hole solution.  Sec.~\ref{sec:interior}  provides a metric, alternative to the isotropic one, which allows exploration of the black hole interior as well as two other sectors which may be interpreted, respectively, as a spacelike naked singularity in asymptotically flat spacetime, and a world lying between two timelike singularities.  In Sec.~\ref{sec:dilaton} we show that our solution corresponds to the extended dilatonic black hole family.  

Sec.~\ref{sec:par} connects the mass and charge of a varying $\alpha$ black hole with the formal parameters of the solution.  Sec.~\ref{sec:extreme} shows that the framework provides no exact counterpart to extremal RN black holes, but that \emph{nearly extremal} black holes have very special properties. We discuss the geometry of externally imposed electromagnetic fields in the background geometry of a nearly extremal black hole.  The thermodynamic functions of a varying $\alpha$ black hole are found \emph{ab initio} in Sec.~\ref{sec:thermo}; we give a trick that considerably simplifies the calculation of the electric potential.  We remark, in agreement with earlier opinions, that cosmological $\alpha$ growth does not compromise the second law; likewise, it cannot drive a charged black hole to become a naked singularity.  In the appendix  we derive directly the solution and properties of the magnetically charged black hole in the framework; they coincide with those one would expect from duality considerations. 

In what follows we take the metric signature as $\{-1,0,0,0\}$ and denote
the temporal coordinate by $t$ and the others by $x^i$ with
$i=1,2,3$.  Greek indices run from 0 to 3.    
 
 \section{Equations and boundary conditions}\label{sec:equations}
 
From the action follow the equations relevant for a bare
black hole (it is convenient to subtract the trace part of Einstein's equations):
\begin{eqnarray}
0&=&(e^{-2\psi}F^{\mu\nu})_{;\nu}
\label{Maxwell}
\\
\psi_{,\mu;}{}^\mu &=&-{\kappa^2\over
2}e^{-2\psi}F_{\mu\nu}F^{\mu\nu}
\label{wave_equation}
\\
R_{\mu\nu}&=&{2 G\over c^4}\Big[e^{-2\psi}
\Big(F_{\mu}{}^\alpha F_{\nu\alpha}-{\scriptstyle 1\over
\scriptstyle 4} g_{\mu\nu}
F^{\alpha\beta}F_{\alpha\beta}\Big)
\nonumber
\\
&+&{1\over\kappa^2}\psi_{,\mu}
\psi_{,\nu}\Big].
\label{Einstein}
\end{eqnarray}
Eq.~(\ref{Maxwell}) shows that whereas $e^{-2\psi}$ is the vacuum electrical permittivity, $e^{2\psi}$ plays the role of vacuum magnetic permeability.   For in 3-d dimensional language appropriate to an inertial frame with 4-velocity $\partial/\partial t$ and Cartesian spatial coordinates, it  reads $\vec\nabla\cdot(e^{-2\psi}  \vec E)=0$ and $\vec\nabla\times (e^{-2\psi} \vec B) =c^{-1}\partial (e^{-2\psi}\vec E)/\partial t$.  Accordingly, the speed of light $c$ is constant in the framework; we shall henceforth set $c=1$.

Eqs.~(\ref{Maxwell})-(\ref{Einstein}) must be supplemented by suitable boundary conditions. 
Asymptotically we must require that the geometry approach
Minkowski's, and that $F^{\mu\nu}\rightarrow 0$ while
$\psi\to\psi_c= \textrm{const.}$, physically the coeval value of $\psi$ in the cosmological model in which our solution is embedded.  At the putative black hole horizon
$\mathcal{H}$ we must require that all physical quantities,
such as the curvature (Ricci) scalar $R=g^{\mu\nu}R_{\mu\nu}$  as well as $F_{\mu\nu}F^{\mu\nu}$
be bounded (otherwise $\mathcal{H}$ would be a singularity). Now from the trace
of Einstein's equation (\ref{Einstein}) we get
\beq
R={2 G\over \kappa^2 }(\psi^{,\mu}
\psi_{,\mu}).
\label{scalar_c}
\eeq
Regularity of $R$  at $\mathcal{H}$ thus requires that $\psi^{,\mu}
\psi_{,\mu}$ be bounded there.

It is easy to check that the Schwarzschild and Kerr black holes
are exact solutions of
Eqs.~(\ref{Maxwell})-(\ref{Einstein}) with
$\psi=\textrm{const.}$ and $ F^{\mu\nu}=0$.  On the other
hand, the RN black hole with
$\psi=\psi_c$ is not a solution because the r.h.s.
term of Eq.~(\ref{wave_equation}) prevents derivatives of
$\psi$ from vanishing ($F_{\mu\nu}F^{\mu\nu}\neq 0$ for
RN). 

We now assume a static spherically symmetric situation.  In isotropic coordinates ($\{x^1,x^2,x^3\}=\{r,\theta,\varphi\})$ the metric is
\beq 
ds^2=-e^{2A(r)} dt^2+e^{2B(r)} (dr^2
+r^2 d\Omega^2),
\label{metric}
\eeq
where $d\Omega^2\equiv d\theta^2+\sin^2\theta\, d\varphi^2$. 
The boundary conditions require that $A(r)\to 0$ and $B(r)\to 0$ as $r\to \infty$.
Assuming there is only electric charge means that the only component of $F^{\mu\nu}$ is $F^{01}=F^{tr}$, which will be a function of $r$ only. Also $\psi=\psi(r)$.  Hence \Eq{Maxwell} gives
\beq
F^{tr}=Q{e^{2\psi-A-3B}\over r^2},
\label{Field}
\eeq
with $Q$ an integration constant.  Asymptotically we expect to recover a radial Coulomb field.  We thus identify $Q e^{2\psi_c}$ as the electric charge measured {\it a\ la} Gauss from infinity.

From the last result we may compute 
\beq
F^{\alpha\beta}F_{\alpha\beta} =-2Q^2 {e^{4\psi-4B}\over r^4}. 
\label{F^2}
\eeq
 Substituting this in  \Eq{wave_equation} we have the scalar equation
\beq
(e^{A+B} r^2 \psi')'=\kappa^2 Q^2 {e^{2\psi+A-B}\over r^2}.
\label{sc2}
\eeq
where here and henceforth `` $'$ " denotes the derivative with respect to $r$.

We now make \Eq{Einstein} explicit:
\bea
tt:\qquad\ A'' +{2A'\over r}+ A'B'+A'^2 = H\qquad\qquad\qquad&&
\label{Ett}
\\
rr:\ A''+2B''+{2B' \over r}-A'B'+A'^2= H-{2G\over \kappa^2}\psi'^2\quad&&
\label{Err}
\\
\theta\theta:\qquad  B''+ A'B'+B'^2+{A'\over r}+{3B' \over r}=- H\quad\qquad&&
\label{Ethth}
\\
H\equiv GQ^2{e^{2\psi-2B}\over r^4}\qquad\qquad\qquad\qquad&&
\eea

\section{Spherically symmetric static solutions}\label{sec:BH}

\subsection{Excluding non black holes}\label{sec:chara}

There  are actually several families of solutions of Eqs.~(\ref{sc2})-(\ref{Ethth}).  We proceed to exclude the irrelevant ones.

We define the variables $C=A+B$ and $D=A-B$. Adding \Eqs{Ett}{Ethth} gives
\beq
C'^2+C''+3\frac{C'}{r}=0,
\eeq
which can be integrated to
\beq
\big(r^3 (e^C)'\big)'=0.
\eeq
Integrating once more and demanding that $C(r)\to 0$ as $r\to\infty$ (asymptotic flatness) leads to
\beq
e^C= 1 -\frac{\varepsilon b^2}{r^2},
\label{C}
\eeq
where $\varepsilon b^2$ stands for an integration constant.  We should thus consider the choices $\varepsilon=0\  {\rm or} \pm 1$.  The $b$ will be taken as positive (no loss of generality as will become clear in \ref{sec:e=1}); its dimension is that of length.

Now $\mathcal{A}(r)=4\pi e^{2B} r^2$ is the area of a 2D-section of space concentric with the origin of the coordinates.  With help of \Eq{C} this may be rewritten as
\beq
\mathcal{A}(r)=4\pi  r^2 \big(1 -\frac{\varepsilon b^2}{r^2}\big)^2 e^{-2A}.
\label{Area}
\eeq
The horizon $\mathcal{H}$ must be a null
 and spherically symmetric hypersurface, so it corresponds to one of those sections and is represented by the
equation $r-r_\mathcal{H}=0$ with $r_\mathcal{H}$ a
constant.    Further, the Killing vector $\xi^\mu$
corresponding to stationarity is $\{1,0,0,0\}$ with norm
$g_{tt}=-e^{2A}$.  This vector must become null on the horizon, for
otherwise $\mathcal{H}$'s generator would not be in a
symmetry direction.  Hence $e^{2A}\to 0$ on $\mathcal{H}$.  Thus the last factor in $\mathcal{A}(r)$
diverges at $\mathcal{H}$.

However, $\mathcal{A}(r_\mathcal{H})$ cannot be infinite, for if it were so, $\mathcal{A}(r)$ would be decreasing with $r$ just outside $\mathcal{H}$.  One consequence would be the existence of a spherically symmetric outgoing congruence of null rays, launched from just outside $\mathcal{H}$, with initially decreasing area, that is with initial positive convergence.  But according to the focusing theorem, in the presence of fields obeying the null positive energy condition [which holds for action (\ref{action})], such congruence must reach a caustic (singularity) in a finite stretch of affine parameter.  Of course a spherically symmetric singularity outside $\mathcal{H}$ would negate its horizon character.    

Thus we must require $\mathcal{A}(r_\mathcal{H})<\infty$. We conclude that the product  of the first factors  in \Eq{Area} must cancel the divergence of $e^{-2A}$.  Now for  $\varepsilon=-1$ such product can vanish only as $r\to \infty$, but the horizon cannot lie at infinity.  Thus there can be no black hole solution for $\varepsilon=-1$.  We proceed to show that assuming $\varepsilon=0$ also fails to produce black holes.

For $\varepsilon=0$ we must have $e^C=1$ and so $B=-A$. 
The divergence of $\mathcal{A}$  is defused if $r_\mathcal{H}=0$ and if, for $r\to 0$, $e^{2A}$ vanishes as $r^2$  or slower.  In the second case we would have $\mathcal{A}\to 0$ as $r\to 0$.  This is unacceptable since a zero area surface cannot be traversed by a finite sized object, so that were the horizon to have zero area,  the black hole's interior could not be accessed by real particles.  This does not correspond to the usual notion of black hole. Thus for $\varepsilon=0$ only the behavior $e^{2A}\propto r^2$ as $r\to 0$ is acceptable.

Let us subtract \Eq{Ethth} from \Eq{Ett}:  
\beq
A''+\frac{2A'}{r}=GQ^2\frac{e^{2\psi+2A}}{r^4}.
\label{A''}
\eeq
Now since as $r\to 0$, $A\sim \ln r$, we deduce from this equation that $e^\psi$ must remain bounded and nonvanishing at $\mathcal{H}$ so that $|\psi|$ cannot blow up there. It follows that $r\psi'\to 0$  as $r\to 0$.  This would imply that the l.h.s. of \Eq{A''} with $A=-B$ must vanish at the horizon.  However, with $e^{2A}\propto r^2$ and $e^{2\psi}$ nonvanishing the r.h.s. is nonvanishing there.  The contradiction can be traced to the assumption that $\varepsilon=0$, which we must thus reject.  The only $\varepsilon$ which is consistent with black holes is thus $\varepsilon=1$.

\subsection{Black hole solutions\label{sec:e=1}} 

 For $\varepsilon=1$ the divergence of $\mathcal{A}$ is prevented  if $r_\mathcal{H}=b\neq 0$ and $e^{2A}$ vanishes for $r\to b$ as $(r-b)^2$ or slower.  Again exclusion of zero horizon area leaves us only the first option.  In this case by \Eq{C} $e^{-2B}$ is bounded at $\mathcal{H}$.  Thus the requirement of bounded $F^{\alpha\beta}F_{\alpha\beta}$ tells us, again, that $\psi$ must not tend to $+\infty$ as $r\to b$.  Furthermore, according to \Eq{scalar_c} $\psi'$ must remain bounded as $r\to b$.  This rules out behavior like $\psi\to -\infty$ at $\mathcal{H}$.  

Subtracting \Eqs{Ett}{Ethth},  rewriting $A=\half(C+D)$ and $B=\half(C-D)$ and replacing the source term by means of \Eq{sc2} gives
 \beq
D'C'+D''-\frac{C'}{r}+2\frac{D'}{r}=\frac{2G}{\kappa^2 r^2 e^C}\left(r^2e^{C} \psi'\right)'.
\eeq
Using \Eq{C} we get after some algebra 
\beq
\left(D'(r^2-b^2)\right)' -\frac{2b^2}{r^2}= \frac{2G}{\kappa^2} \left((r^2-b^2)\psi'\right)',
\eeq
which  equation integrates to
\beq
D'(r^2-b^2)+\frac{2b^2}{r}=\frac{2G}{\kappa^2} (r^2-b^2)\psi' + a b,
\label{inte2}
\eeq
where $a$ is a new (dimensionless) integration constant. Let us now introduce the new radial coordinate
\beq
\xi=\int \frac{dr}{r^2-b^2} =\frac{1}{2b}\ln\left(\frac{r-b}{r+b}\right).
\label{xi}
\eeq
We note that $\xi\to -\infty$ at $\mathcal{H}$ and $\xi\to 0$ at spatial infinity.
Dividing \Eq{inte2} by $r^2-b^2$, going over to variable $\xi$, and performing an additional integration we obtain
\beq
D + \ln\left(1-\frac{b^2}{r^2}\right)-a b \xi = \frac{2G}{\kappa^2}(\psi-\psi_c),
\label{dpsi}
\eeq
where the new integration constant has been chosen so as to enforce the asymptotic conditions $\psi\to \psi_c$ and $D\rightarrow 0$ as $r\to +\infty$.

Now the first two terms in \Eq{dpsi} add up to $2A$ which, we have seen, must behave as $2\ln(r-b)$ as $r\to b$.  Since $\psi$ is bounded there, the term $-a b \xi$ must diverge there as $-2\ln(r-b)$, which is possible only if $a=4$.  Thus \Eq{dpsi} can be written as
\beq
2A=2\ln \left(\frac{r-b}{r+b}\right) +{2G\over \kappa^2}(\psi-\psi_c),
\label{Aint}
\eeq
so that the only unknown now is $\psi$.

We  now focus on the scalar equation (\ref{sc2}) where we substitute $e^{2A}$ from the above results.
In terms of the new field variable  
\beq
u\equiv 2(1+G/\kappa^2)(\psi-\psi_c) +4 b\xi
\label{u}
\eeq 
(whose range is exactly that of $\xi$) it becomes
\beq
{d^2u\over d\xi^2} = q^2 e^u;\qquad q^2\equiv 2(G+\kappa^2)Q^2 e^{2\psi_c}. 
\label{q} 
\eeq
This is the equation of motion in time $\xi$ of a particle moving in an exponential potential.  Its first integral is
\beq
{1\over 2}\left({du\over d\xi}\right)^2=E+q^2 e^u,
\eeq
where $E$ is the analog of the particle's conserved energy.

Since $\psi'$ remains bounded as $r\to b$,  $d\psi/d\xi\to 0$ and $du/d\xi\to 4b$ at $\mathcal{H}$.  In addition $u\to -\infty$, so the $q^2 e^u$ term vanishes.  Thus the above equation requires that $E=8b^2$.  

Since $u\to 0$ as $r\to +\infty$, and the above equation has no turning point, it is clear thats $u$ increases with $\xi$.
Thus the integral of the above equation is
\beq
\int{du\over \sqrt{8b^2+q^2 e^u}} =\surd 2  \int\  d\xi
\label{intp}
\eeq
with positive root. 

The integration is done as follows.  In terms of the variable $v\equiv u-\ln (8b^2/q^2)$ the integral of \Eq{intp} with the correct boundary conditions included is
\beq
4b\xi=\int_{\ln(q^2/8b^2)}^0 {dv'\over \sqrt{1+e^{v'}}}\,.
\eeq
Now substitute $e^v=\sinh^2 x$ to obtain the integral of $2{\rm csch}\, x$ over $x$ which is $2\ln[{\rm csch}\, x (\cosh x-1)]$.   Simplifying with help of identities for hyperbolic functions, and reverting to variable $u$ one obtains
\bea
4b\xi&=& \ln\Sigma\big((q^2/8b^2) e^u\big)-\ln\sigma;
\\
\Sigma(z)&\equiv& 1+2 z^{-1} - 2 z^{-1} \sqrt{1+z}\ ,
\label{Sigma}
\\
\sigma&\equiv& \Sigma(q^2/8b^2),
\label{sigma}
\eea
where the square root is expressly the positive one.

It follows from \Eq{sigma} that
\beq
(32b^2/q^2)\sigma=(1-\sigma)^2,
\label{iden}
\eeq
It is clear from this that $\sigma$ cannot be negative.  Further,  \Eq{sigma} shows that  $0<\sigma<1$.

To recover $\psi(\xi)$ first solve \Eq{Sigma} for $z$:
\beq
z= 4 \Sigma(z) \big(1- \Sigma(z)\big)^{-2}\,.
\eeq
In this last substitute on the l.h.s. $z\Longrightarrow (q^2/8b^2) e^u$ and on the r.h.s. $\Sigma(z)\Longrightarrow \exp(4b\xi+\ln\sigma)$,  and invoke \Eq{xi} to get
\beq
e^u=\sigma\left({32b^2\over q^2}\right)\left({r-b\over r+b}\right)^2 \left(1-\sigma\left({r-b\over r+b}\right)^2\right)^{-2}.
\eeq
Now take the logarithm and substitute $u$ here from \Eq{u}; after canceling a term and taking \Eq{iden} into account we get
\beq
\psi=\psi_c+{\kappa^2\over G+\kappa^2}\ln
\left(\frac{1-\sigma}{1-\sigma\big({r-b\over r+b}  \big)^2  }\right).
\label{psi}
\eeq

At this point recall \Eq{Aint}; Exponentiating it and substituting from our last result gives
\beq
e^{2A}=\left({r-b\over r+b}  \right)^2 \left({1-\sigma\over 1-\sigma\big({r-b\over r+b}  \big)^2 }\right)^{2G\over G+\kappa^2}
\label{e2A}
\eeq
Further, from $B=C-A$ and \Eq{C} follows
\beq
e^{2B}=\left( 1+{b\over r} \right)^4 \left({ 1-\sigma\big({r-b\over r+b}\big)^2\over 1-\sigma }\right)^{2G\over G+\kappa^2}
\label{e2B}
\eeq

We have so far supposed that $b>0$.  Are values $b\leq 0$ permitted too?
Note that $b=0$ amounts to taking $\varepsilon=0$ in \Eq{C} and that case was already ruled out there.  As for $b<0$, suppose we formally switch the sign of $b$ in our solution (\ref{psi})-(\ref{e2B}).  It is then clear that $e^{2B}=0$ and $e^{2A}=\infty$ on the surface $r=|b|$, meaning that surface has zero area and is an infinite blueshift surface.  Thus no physical body could penetrate from $r>|b|$ while photons launched outwardly from near the surface would reach distant observers with arbitrary large energy.  The object in question would obviously be pathological. This means $b$ must be positive, as assumed.

Eqs.~(\ref{psi})-(\ref{e2B}) specify the unique exact solution for a nonrotating charged black hole in general relativity with varying $\alpha$ electrodynamics.  Several special cases are of interest.  

\begin{itemize}

\item
For $Q\to 0, \sigma\to 0$, and the solution takes the form of a Schwarzschild metric in isotropic coordinates with $\psi=\psi_c$.  As mentioned in Sec.~\ref{sec:equations}, the Schwarzschild solution should be a solution also in the varying $\alpha$ theory.

\item 
For $\kappa\to 0$, $\psi\to \psi_c$ and both exponents in the last factors in \Eqs{e2A}{e2B} become 2.  Thus
\bea
e^{2A}&=&\left({r^2-b^2\over r^2+b^2+ 2 b\big({1+\sigma\over 1-\sigma}\big)r}  \right)^2
\\
e^{2B}&=&{1\over r^4}\left( r^2+b^2+ 2 b\Big({1+\sigma\over 1-\sigma}\Big)r  \right)^2
\eea
This is precisely the RN metric in isotropic coordinates (see for example Ref.~\onlinecite{EvaBek}).  Anticipating \Eqs{M}{b1} we find the parameter correspondence to be $b=\fourth (R_+-R_-)$  and $\sigma=R_-/R_+$ with $R_\pm= GM\pm \sqrt{G^2 M^2 - G Q^2 e^{2\psi_c}}$ and $M$ the mass.

\end{itemize}

\section{Regions and sectors of the solution}\label{sec:interior}

It is worthwhile noting that under the change of variable $r\hookrightarrow \rho \equiv b^2/r$, the metric (\ref{metric}) with \Eqs{e2A}{e2B} goes over into itself with $\rho$ everywhere replacing $r$.  Thus the interval $r\in (0,b)$ is mapped onto the black hole exterior $r\in (b,\infty)$ with the horizon being a fixed point.  Just as in the case of the RN solution in isotropic coordinates, the  the metric (\ref{metric}) with \Eqs{e2A}{e2B} covers the black hole exterior twice, but does \emph{not} cover any part of the black hole interior.

In the RN case this sort of problem is resolved by passing to Schwarzschild style coordinates (squared radial coordinate gives area).  Such a transformation is intractable  here as it entails solution of a higher order algebraic equation.  Because of this we opt for a slightly different radial coordinate 
\beq
\varpi\equiv \left(\frac{r-b}{r+b}\right)^2.
\eeq

This can be inverted to get 
\beq
r=b\frac{1+\surd\varpi}{1-\surd\varpi}\ .
\eeq
Actually a second solution may be obtained by switching the signs of the radicals.  This doubling corresponds to the two radial coordinates $r$ and $b^2/r$ which cover the same (exterior) region, as discussed earlier in this section.

Transforming metric (\ref{metric}) with \Eqs{e2A}{e2B} from $r$ to $\varpi$ we have
\bea
&&ds^2=-\varpi\left(\frac{1-\sigma}{1-\sigma \varpi}\right)^{2G\over G+\kappa^2}dt^2
\label{metric'}
\\
&+&\frac{16b^2}{(1-\varpi)^2}\left(\frac{1-\sigma \varpi}{1-\sigma}\right)^{2G\over G+\kappa^2}\left(  \frac{d\varpi^2}{\varpi(1-\varpi)^2}+d\Omega^2\right). \nonumber
\eea 
Likewise writing down the field $\psi$ from \Eq{psi} we have
\beq
\psi=\psi_c+{\kappa^2\over G+\kappa^2}\ln
\left(\frac{1-\sigma}{1-\sigma\varpi  }\right).
\label{psi1}
\eeq
It should be clear that the new coordinates are suitable so long as $\sigma<1$; the case $\sigma\to 1$ is discussed in Sec.~\ref{sec:extreme}.

The black hole exterior is covered by the coordinate domain $\varpi\in (0,1)$ with $\varpi=1$ being spatial infinity and $\varpi=0$ being the event horizon $(r=b)$.  The change of coordinate has now put the horizon's interior in view: it is the domain $\varpi\in (-\infty, 0)$ wherein the $t$ coordinate becomes spacelike and $\varpi$ timelike.   

We identify $\varpi=-\infty$ as the central singularity.  The reasoning is as follows.  From \Eqs{metric'}{psi1} and comparing with \Eq{scalar_c} we have
\beq
R\propto \psi_{,\mu}\psi^{,\mu}={\varpi(1-\varpi^4)}{\big(1-\sigma\varpi\big)^{-\frac{4G+2\kappa^2}{G+\kappa^2}}} .
\label{R1}
\eeq
Thus the scalar curvature diverges for $\varpi\to -\infty$ signifying that $\varpi=-\infty$ is a true singularity. Since it borders a region where $g_{\varpi\varpi}<0$ this singularity is spacelike (normal with negative norm).  Further, from the area of a $\varpi=$ const. surface,
\beq
\mathcal{A}(\varpi)\propto (1-\sigma\varpi\big)^{\frac{2G}{G+\kappa^2}}  (1-\varpi)^{-2},
\label{area'}
\eeq
we observe that the singularity has vanishing area.  Thus $\varpi=-\infty$ is a central singularity, in all respects like the one in the Schwarzschild solution, and in contrast to the timelike singularity of the RN solution. 

For $\kappa\neq 0$ there is no second (inner or Cauchy) horizon, such as we have in the RN solution.  It is true that $g_{tt}$ can vanish not only at $\varpi=0$ but,  provided $\kappa^2<G$, also at $\varpi=\infty$.   However, as clear from \Eq{R1}, $\varpi=\infty$ is, like $\varpi=-\infty$, a point of unbounded curvature, and cannot be a horizon.  Thus we are left with a single horizon, $\varpi=0$.

But then how does the RN solution (for $\kappa=0$) manage to have two horizons?  For this Maxwellian electrodynamics case the scalar $R$ vanishes identically since $\psi$ must be constant (see \Eqs{psi}{R1}).  There is then no longer any reason for $\varpi=\infty$ to be a singularity, and thus it might be a horizon.  To see that it is introduce the area radial coordinate
\beq
\varrho=\frac{4b}{1-\sigma} \frac{1-\sigma \varpi}{1-\varpi},
\label{varpi}
\eeq
for which $\mathcal{A}(\varrho)=4\pi \varrho^2$ just as with the  radial coordinate in the usual form of Schwarzschild's metric.
In terms of $\varrho$ we have
\beq
g_{tt}=-\frac{\big((1-\sigma)\varrho-4b\big)\big((1-\sigma)\varrho-4b\sigma\big) }{ (1-\sigma)^2\varrho^2 }.
\eeq
It is now clear that there two horizons, at $\varrho=R_\pm$, with $R_-/R_+=\sigma$ and $\fourth (R_+-R_-)=b$.  These last are precisely the relations quoted for RN at the end of Sec.~\ref{sec:e=1}.
In agreement with our previous remarks we note that, according to \Eq{varpi}, the inner horizon $R=R_-$ corresponds to $\varpi=-\infty$.
Thus RN is the only charged spherical black hole sporting an inner horizon.  More on this, in light of GHS, in Sec.~\ref{sec:dilaton}.  

Metric (\ref{metric'}) describes not only black holes, but also other denizens of the gravitational world.  

In the domain $\varpi\in (1,1/\sigma)$,  $t$ is a timelike coordinate.  The boundary  $(\varpi=1)$ is spatial infinity as can be seen because $\mathcal{A}\to\infty$ as $\varpi\to 1_+$ with $R$ vanishing there.  And because $R$ diverges as $\varpi\to 1/\sigma$, this second boundary is also a singularity which is timelike in character because $g_{\varpi\varpi}>0$ as one approaches it  from lower $\varpi$.  It can be shown from \Eq{area'} that  $d\mathcal{A}/d\varpi<0$ throughout the domain under discussion: the area of 2-surfaces increases monotonically with $\varpi$ as we go from singularity to spatial infinity.    Thus  $\varpi\in (1,1/\sigma)$ spans the static asymptotically flat spacetime around a naked timelike singularity.  This spacetime is obviously distinct from the black hole one. 

Finally we study the domain $\varpi\in (1/\sigma,\infty)$.  Metric (\ref{metric'}) is real there only when $2G/(G+\kappa^2)$ is a rational number which, when maximally reduced,  is either of form $(2n_1+1)/(2n_2+1)$ or $2n_1/(2n_2+1)$ with $n_1, n_2$ two integers.  However, in the first case   $t$ becomes spacelike in the said domain while $\varpi$, $\theta$ and $\varphi$ all become timelike.  The metric's signature is thus unphysical and we must reject any physical interpretation of this case.

By contrast, for  $2G/(G+\kappa^2)=2n_1/(2n_2+1)$ the metric's signature is the usual one, with $t$ a timelike coordinate in the whole domain $\varpi\in (1/\sigma,\infty)$.  According to \Eq{R1} the scalar curvature diverges both as $\varpi\to 1/\sigma$ and $\varpi\to\infty$ from within the domain, so that the said boundaries are timelike singularities.  
 The spacelike distance between them,
 \beq
\ell=\frac{4b}{(1-\sigma)^\frac{G}{G+\kappa^2} }\int_{1/\sigma}^\infty \frac{(1-\sigma \varpi)^\frac{G}{G+\kappa^2}d\varpi}{ \surd \varpi\ (1-\varpi)^2 },
\eeq
 converges at both limits of the integral, so we may speak of a finite static spacetime lying between two spherically symmetric timelike singularities of vanishing area.    This spacetime is evidently not asymptotically flat.  
 
 The two singularities are point charges of clearly opposite signs and the same magnitude, since by Gauss' law the electric flux lines issuing from one must, by dint of the symmetry, end up in the other.  How come the two charges do not pull each other together?  A simple calculation shows that a freely falling particle will oscillate between the two singularities, and find it impossible to approach either, no matter how large its conserved energy is.  Gravity is thus repulsive in nature in the said region, and this must be the agent that balances the charges' attraction.

\section{Identity with dilatonic black holes}\label{sec:dilaton}

We now demonstrate the identity of our solution with the spherical dilatonic black holes by comparing with GHS's formulation~\cite{ghs}. GHS use a radial coordinate $\mathcal{R}$ such that $g_{tt}=-1/g_\mathcal{RR}$. To convert metric~(\ref{metric'}) to this form we obviously have to require
\beq
\frac{4 b\, d\varpi}{(1-\varpi)^2}=\pm d\mathcal{R}.
\eeq
Taking the positive sign so that $\mathcal{R}$ increases with $\varpi$, and choosing the integration constant (the zero of $\mathcal{R}$) with hindsight we get
\beq
\mathcal{R}=\frac{4b}{1-\varpi}+\frac{4b\sigma}{1-\sigma}\, .
\eeq
Inverting this we can put metric~(\ref{metric'}) in the form
\beq
ds^2=-\lambda^2 dt^2 + \frac{d\mathcal{R}^2}{\lambda^{2}} + \mathcal{R}^2\Big(1-\frac{4b\sigma}{(1-\sigma)\mathcal{R}}\Big)^{\frac{2G}{G+\kappa^2}}d\Omega^2,
\label{metric''}
\eeq
with
\beq
\lambda^2\equiv \Big(1-\frac{4b}{(1-\sigma)\mathcal{R}}\Big) \Big(1-\frac{4b\sigma}{(1-\sigma)\mathcal{R}}\Big)^{\frac{G-\kappa^2}{G+\kappa^2}}.
\label{lambda2}
\eeq
We may also translate $\psi$ from \Eq{psi1} to the form
\beq
\psi=\psi_c+\frac{\kappa^2}{G+\kappa^2}\ln \Big(1-\frac{4b\sigma}{(1-\sigma)\mathcal{R}}\Big).
\label{psi2}
\eeq

Strictly speaking dilaton theory is based on equations (\ref{Maxwell})-(\ref{Einstein}) but with the choice $\kappa^2=G$~\cite{ghs}.  To investigate stability of the dilatonic black holes with respect to changes of the parameters of the theory, GHS also considered the case of generic $\kappa$ (which they denote $a$) but setting $\psi_c=0$. Our \Eqs{metric''}{lambda2} agree in form with
theirs; the difference between their expression for the dilaton field and our \Eq{psi2} is immediately understood if we recall that our $\psi/\kappa$ corresponds to GHS's dilaton, and that GHS deal with the magnetic charge case,  c.f. \Eq{psi*} of the appendix below. 

GHS remark that the zero of $\lambda^2$ at $\mathcal{R}=4b\sigma/(1-\sigma)$ is really a singularity, except in the case $\kappa=0$ in which it marks the inner horizon of the RN geometry.  This tallies with the point made in Sec.~\ref{sec:interior} that $\varpi=-\infty$ is the central singularity in the generic case.  Together with Gibbons and Maeda~\cite{GM} and GHS~\cite{ghs} we conclude that the inner horizon becomes unstable and metamorphoses into a singularity as $\kappa$ departs from zero.

We shall connect the two black hole parameters to observables in a different way than did GHS.    For this purpose we revert to metric (\ref{metric}) with \Eqs{e2A}{e2B}.

\section{Physical observables of black holes with varying $\alpha$}\label{sec:par}

How do the black hole parameters $b$ and $Q$ relate to the observable properties of the black hole for generic $\kappa$?  In \Eq{Field} the radial electric field as $r\to\infty$ is $Q e^{2\psi_c}/ r^2$.  By asymptotic flatness the $4\pi r^2$ is, in this limit,  a good measure of the area of a sphere concentric with the black hole, so that we must conclude that the charge, as inferred from Gauss' law, is $Q e^{2\psi_c}$.  This is the observable charge (as seen from infinity).  We refer the reader to the discussion in Ref.~\onlinecite{bek} to the effect that  $Q$, the strength parameter in the electric current, is  rather the {\em conserved} charge.  In agreement with this, the observed charge at infinity would undergo evolution in an expanding universe (where $\psi_c$ would be given by the time dependent solution to the cosmological problem).

To obtain the observable mass $M$ we expand \Eq{e2A} in powers of $1/r$:
\beq
e^{2A}=1-{4b\over r} \left( 1+{2G\sigma\over (G+\kappa^2)(1-\sigma)} \right) +\mathcal{O}(r^{-2})
\eeq
Because asymptotically $r$ is the radius, one can identify the coefficient of $-1/r$ here as $2GM$.
Replacing the $\sigma$ in terms of $(1-\sigma)^2$ by means of \Eq{iden}, and substituting $\sigma$'s 
value from \Eq{sigma} we finally get the observable mass,
\beq
M=2 b \left({1\over G} +{\sqrt{1+\fourth(G+\kappa^2) b^{-2} Q^2 e^{2\psi_c}}-1\over  (G+\kappa^2)}\right)
\label{M}
\eeq
where, again, the positive square root is chosen.

Solving this expression for $b$ in terms of $M$ and $Q$ by first manipulating it into the form of a quadratic in $b$ we get, for $\kappa^2\neq G$,
\beq
b={-G\kappa^2 M+ \sqrt{G^4 M^2 - (G-\kappa^2)G^2 Q^2 e^{2\psi_c}}\over 2(G-\kappa^2)}
\label{b1}
\eeq
The second solution to the quadratic---corresponding to a negative signed radical---turns out to be extraneous.  It would be the solution to \Eq{M} if the radical in the latter were negative: the solution procedure described above entails squaring that radical, and so adds an unphysical solution which must be rejected by hand.

By contrast, if $\kappa^2=G$ we naturally get just one solution:
\beq
b=\half GM-\fourth Q^2 e^{2\psi_c}/M.
\label{b}
\eeq
To avoid negative $b$, which would be meaningless for an horizon, one demands
$
|Q|\leq \sqrt{2 G} M e^{-\psi_c}.
$

Thus for any $\kappa^2$, the ``no hair'' principle is satisfied: for any pair $\{M, Q\}$ there is just one black hole, specified by \Eqs{e2A}{e2B} with the parameters given by \Eq{b1} or \Eq{b} together with \Eq{sigma}.  We do not count90 $e^{\psi_c}$ as a black hole parameter since it is set by the cosmological model in which our solution is to be embedded.

Whether for $\kappa^2>G$ (strongly coupled $\alpha$ variability) or  $\kappa^2<G$ (weakly coupled varying $\alpha$ theory)  \Eq{b1} gives nonnegative $b$ only for
$
|Q|\leq  \sqrt{G+\kappa^2}\,M e^{-\psi_c}.
$
Thus for all values of $\kappa^2$, charged black holes can be had only for
\beq
|Q|\leq  \sqrt{G+\kappa^2}\,M e^{-\psi_c}.
\label{bound}
\eeq

For  fixed $M$ the black hole family is a one-parameter sequence; the parameter can be either $b$ or $Q e^{\psi_c}$ or $\sigma$.  Along the sequence $b$ decreases monotonically with $|Q|e^{\psi_c}$ from $b=GM/2$ at $Q=0$ towards its zero point at $|Q|= \sqrt{G+\kappa^2}\,M e^{-\psi_c}$, in the vicinity of  which $b\propto  (\sqrt{G+\kappa^2}-|Q|e^{\psi_c})$. According to \Eq{iden} $\sigma\propto Q^2 e^{2\psi_c} $ as $|Q|\to 0$ while $\sigma\to 1$ for $b\to 0$. 

From \Eq{e2B} with $r=b$  we have for the horizon area 
\beq
\mathcal{A}(b)= {64\pi b^2\over G\hbar (1-\sigma)^{2G/(G+\kappa^2)}}\,.  \label{newarea}
\eeq
Numerically it is found that $\mathcal{A}(b)$ decreases monotonically with $|Q|e^{\psi_c}$ at fixed $M$.  As evident from \Eq{iden}, $\mathcal{A}(b)$ tends to zero as $|Q|\to \sqrt{G+\kappa^2}M e^{-\psi_c}$ and $\sigma\to 1$.  As $|Q|\to 0$, $\mathcal{A}(b)$ tends to its Schwarzschild value $16\pi G^2M^2$.

\section{Nearly extremal black holes}\label{sec:extreme}

In Einstein-Maxwell theory, for which the charged black holes are RN, the extremal black holes are those for which $Q$ attains its least upper bound $\surd G M$.  In our framework extremality would correspond to the saturation of inequality~(\ref{bound}).   We have just seen that this corresponds to $b\to 0$ and $\sigma\to 1$ in which limit the horizon area vanishes.  But as argued in Sec.~\ref{sec:chara}, an object with zero horizon area cannot be regarded as a black hole.  Hence, in the framework there is no exactly extremal black hole (for $\kappa\neq 0$).  This agrees with our rejection of solutions with $\varepsilon=0$ in Sec.~\ref{sec:equations}; $\varepsilon=0$ is equivalent to $b=0$.

In what follows we shall be concerned with the \emph{nearly extremal} black holes in the framework.  Many of their properties can be ascertained most easily by developing the generic formulae in Taylor series in $b$ while setting $(1-\sigma)=4\surd 2\, b/q$ in accordance with \Eq{iden}. 
For example, in the extremal limit
\bea
&&e^{2A}=\left(1+\frac{(G+\kappa^2)M}{r}\right)^{-2G/(G+\kappa^2)}+\mathcal{O}(b),
\label{e2A*}
\\
&&e^{2B} = e^{-2A},
\\
&&e^{\psi}= e^{\psi_c} \left(1+\frac{(G+\kappa^2)M}{r}\right)^{-\kappa^2/(G+\kappa^2)}\kern-5pt+\mathcal{O}(b).
\label{epsi*}
\eea
It is interesting that to leading order $e^\psi$ is just some power of $e^{2A}$.

One interesting application of the above is as follows.  It is known~\cite{LL2} that Maxwellian electrodynamics in a static background geometry can be regarded as electrodynamics in  flat spacetime filled with a medium with electric permittivity and  magnetic permeability both equal to $1/\sqrt{-g_{tt}}$.  But we have remarked that varying $\alpha$ electrodynamics is related to the Maxwellian one by a vacuum permitivitty $e^{-2\psi}$ and a vacuum permeability $e^{2\psi}$.  By compounding the two cases we see that the new electrodynamics in the curved spacetime represented by metric (\ref{metric}) functions like the Maxwellian brand in flat spacetime filled with a medium with permittivity $e^{-A-2\psi}$ and permeability $e^{-A+2\psi}$.  

For the  special case of coupling $\kappa^2=G/2$, we find from \Eqs{e2A*}{epsi*} that the exterior of a nearly extremal electrically charged black hole has unit effective permeability throughout!  This would mean that magnetic lines produced by distant currents would not be bent by the hole's presence (with bending judged with respect to the flat asymptotic space).  And for larger coupling $\kappa^2>G/2$ the effective permeability would be below unity throughout space.  That would make the black hole exterior like a diamagnetic medium which tends to expel magnetic fields.  Therefore, for such strong coupling the black hole would tend to bend external magnetic lines away from itself, much as a superconductor will push out magnetic flux. And a sufficiently strong magnetic field would be able to bodily push the black hole away.

It is easy to see what changes would take place were the black hole charged magnetically.  The basis is explained in the appendix.

\section{Black hole thermodynamics}\label{sec:thermo}

The easiest way to calculate the black hole temperature is via the surface gravity.  At fixed point $x^\mu=\{t, r, 0, 0\}$ has an acceleration vector $a^\mu=\{0, a^r, 0, 0\}$ with $a^r=-\Gamma^t_{rr} (dt/d\tau)^2$.  
For a diagonal metric like that in \Eq{metric} this gives $a^r=e^{-2B} A'$.  Consequently, the invariant acceleration is $\sqrt{a^\mu a_\mu}=e^{-B} A'$.   To these corresponds the local Unruh temperature $T_U=(\hbar/2\pi)e^{-B} A'$.  Redshifting this to infinity we get the global temperature $T_\mathrm{g}=e^A T_U$.  It is reasonable to take $T_{BH}=\lim_{r\to r_\mathcal{H}} T_\mathrm{g}$. Using the specific metric (\ref{metric}) we get in the limit $r\to b$
\beq
T_{BH}=(1-\sigma)^{2G/(G+\kappa^2)} {\hbar\over 16\pi b}.
\label{TBH}
\eeq

An independent approach is afforded by the the Euclidean framework.  According to \Eqs{metric}{e2B} the increment of radial proper length $\ell$ from the horizon to point $r$ is given by $d\ell=e^B dr$ so that
\bea
\ell&=&\int_b^r \left(1+\frac{b}{r}\right)^2 \left({ 1-\sigma\big({r-b\over r+b}\big)^2\over 1-\sigma }\right)^{G\over G+\kappa^2}
\\
&= & 4 (1-\sigma)^{-G/(G+\kappa^2)}(r-b)+\mathcal{O}\big((r-b)^2\big).
\eea
Solving $r-b$ in terms of $\ell$ in \Eq{e2A} gives
\beq
e^{2A}= (1-\sigma)^{4G/(G+\kappa^2)}{\ell^2\over 64 b^2}+\mathcal{O}(\ell^3),
\eeq
so that the Euclidean metric of the $t$-$r$ plane takes the form
\beq
ds^2_{\rm Euc} = (1-\sigma)^{4G/(G+\kappa^2)}{\ell^2\over 64 b^2} d\tau^2 + d\ell^2
\eeq
where $\tau$ is Euclidean ``time''.  If we wish to interpret this $\tau$ as an angle, a conical singularity will occur unless we require that its period be
\beq
\Pi=2\pi \left((1-\sigma)^{4G/(G+\kappa^2)}{\ell^2\over 64 b^2}\right)^{-1/2}.
\eeq
This periodicity is equivalent to a thermal ensemble with temperature $\hbar/\Pi$.  This last is precisely  $T_{BH}$ of \Eq{TBH}.

Yet a third alternative approach is to start with the black hole entropy as a quarter of the horizon area in units of Planck's length squared:
\beq
S_{BH}=\frac{\mathcal{A}(b)}{4G\hbar} ={16\pi b^2\over G\hbar (1-\sigma)^{2G/(G+\kappa^2)}}.
\label{entropy}
\eeq
Then the black hole temperature is
\beq
T_{BH}=\left(\frac{\partial S_{BH}}{\partial M}\right)_Q^{-1}.
\eeq
To evaluate this we should regard $\sigma$ a function of $b$ and $b$ a function of $M$.   Then
\beq
T_{BH}=\frac{G\hbar}{16\pi b}{(1-\sigma)^{2G/(G+\kappa^2)}\over 2\big(\frac{\partial b}{\partial M}\big)_q\left(1+\frac{G b}{G+\kappa^2}\frac{\partial\sigma/\partial b}{1-\sigma}  \right)}
\label{T}
\eeq

The easiest way to obtain $\partial\sigma/\partial b$ is to take the logarithm of \Eq{iden} and differentiate the result with respect to $b$ at fixed $q$:
\beq
\Big(\frac{\partial\sigma}{\partial b}\Big)_q=-\frac{2\sigma(1-\sigma)}{b(1+\sigma)}
\eeq
Therefore, \Eq{T} takes the new form
\beq
T_{BH}=\frac{G\hbar}{16\pi b}\ {(1-\sigma)^{2G/(G+\kappa^2)}\over 2\big(\frac{\partial b}{\partial M}\big)_q\left(1-\frac{G}{G+\kappa^2}\frac{2\sigma}{1+\sigma}\right)}
\eeq
Harmony of this with \Eq{TBH} would entail the identity 
\beq
\Big(\frac{\partial b}{\partial M}\Big)_q\equiv \half G\left(1-\frac{G}{G+\kappa^2}\frac{2\sigma}{1+\sigma}\right)^{-1}.
\label{dbdM}
\eeq
  Although we have been unable to establish this analytically, we have checked it numerically for a large set of values of $Q/M$ and various $\kappa^2/G$.  Thus the Euclidean and thermodynamic calculations of $T_{BH}$ agree. This, by the way, demonstrates that the area formula for $S_{BH}$ does not get corrections from $\alpha$-variability.

As already mentioned in Sec.~\ref{sec:par}, $S_{BH}$ as well as $\mathcal{A}(b)$ vanish in the limit $b\to 0$ in which inequality (\ref{bound}) would be saturated and the black hole would become extremal.  This is in contrast to the situation of the extremal RN black hole which has nonvanishing area and entropy.  The $T_{BH}$ has an even more curious behavior.  As mentioned in Sec.~\ref{sec:extreme} the would be extremal black hole is reached in the limit $b\to 0$ with $(1-\sigma)\sim b$.  It may be seen from \Eq{TBH}   that for $\kappa^2<G$ the temperature vanishes in that limit (just as it does for the RN black hole), while for 
$\kappa^2>G$ it diverges.  As already noticed by GHS, for the true dilatonic black holes ($\kappa^2=G$) $T_{BH}$ remains finite~\cite{ghs}.

In the thermodynamic approach the electric potential of the black hole, $\Phi_{BH}$ is
\beq
\Phi_{BH}=-T_{BH} \left(  \frac{\partial S_{BH}}{\partial (Q e^{2\psi_c}) }\right)_{M} =\left(\frac{\partial M}{\partial (Q e^{2\psi_c})}  \right)_{S_{BH}}\ .
\label{potential*}
  \eeq
since, as discussed in Sec.~\ref{sec:equations},  $Q e^{2\psi_c}$ is the  charge observable from infinity. Since this leads to a very complicated computation, we here determine $\Phi_{BH}$ as the value of the electric potential $-A_t$ in the limit $r\to b$ (provided $A_t$ vanishes asymptotically).  The logic for this prescription is as follows.  The Lagrangian for a charged particle with mass
$\mu$ and (conserved) charge $e$ is
\beq
\mathcal{L}=-\mu\sqrt{-{g}_{\alpha\beta}\frac{dx^\alpha}{d\tau}\frac{dx^\beta}{d\tau}}+e{A}_\alpha\frac{dx^\alpha}{d\tau},
\eeq 
where $A_\alpha$ is the electromagnetic vector potential: $F_{\alpha\beta}=   A_{\beta,\alpha} - A_{\alpha,\beta}$.  In a gauge for which $A_\alpha$ is time independent,  the Lagrangian is also $t$ independent, and we have the conserved canonical momentum 
 \beq
P_t=\frac{\partial\mathcal{L}}{\partial\frac{dt}{d\tau}}=\mu
{g}_{tt}\frac{dt}{d\tau}+e{A}_t.
\eeq  
Since the first term in $P_t$ is \emph{minus} the 
particle's rest plus kinetic energy, $P_t$ is minus the total energy and 
$-e{A}_t$ must be the  particle's
electric energy at the corresponding  point.  The  electric potential of the black hole as measured from infinity can thus be deduced from the value this energy takes on as the particle nears the horizon:
\beq
\Phi_{BH}=-\lim_{r\to b} e^{-2\psi_c} {A}_t.  \label{calib}
\eeq  
Here the factor $e^{-2\psi_c}$ accounts for the fact that the observable charge of the \emph{particle} is $e^{2\psi_c} e$.

Turning to \Eq{Field} we have
\beq
A_{t,r}=e^{2A+2B} F^{tr}= Q{e^{2\psi+A-B}\over r^2}.
\eeq
Substitution from Eqs.~(\ref{psi})-(\ref{e2B}) gives
\beq
A_{t,r}=\frac{Q e^{2\psi_c} (r^2-b^2)}{\Big(r^2+b^2+2\frac{1+\sigma}{1-\sigma} b r\Big)^2}\,.
\eeq
This integrates to
\beq
A_t=-\frac{Q e^{2\psi_c} r}{r^2+b^2+2\frac{1+\sigma}{1-\sigma} b r},
\eeq
which appropriately vanishes as $r\to\infty$.  Now taking the limit $r\to b$ in accordance with \Eq{calib} gives
\beq
\Phi_{BH}=\frac{Q (1-\sigma)}{4b}\,.
\label{potential}
\eeq

We have checked numerically (for a variety of values of $Q/M$ and $\kappa^2/G$) that the Maxwell thermodynamic relation
\beq
\left(\frac{\partial (1/T_{BH}) }{\partial (Q e^{2\psi_c})} \right)_{M} =-\left(\frac{\partial (\Phi_{BH}/T_{BH}) }{\partial M} \right)_{Q e^{2\psi_c}}
\eeq
is indeed obeyed.  Further, we recall that as $\kappa\to 0$ we should recover the RN result.  According to Sec.~\ref{sec:e=1}, for the RN black hole $4b=R_+-R_-$ and $\sigma=R_-/R_+$, where $R_\pm$ are the radii of the two horizons in Schwarzschild-like coordinates.    With these our potential $\Phi_{BH}$ reduces to $Q /R_+$, which is the correct RN result.  Both above checks support the correctness of our result for $\Phi_{BH}$.  Note that $\Phi_{BH}$ depends on $e^{\psi_c}$ through $b$ and $\sigma$.
 
An application of the above is to the issue of whether black hole thermodynamics can usefully constrain the cosmological rate of change of $\alpha$?  Davies, Davis and Lineware (DDL)~\cite{davies} argued from the form of the black hole entropy of a RN black hole that it could not help but decrease if $\alpha$ is increasing as claimed by Webb's group~\cite{webb}. They propose to rule out a cosmologically growing $\alpha$ on the ground it would violate the generalized second law (if one ignores the entropy produced by Hawking emission).  DDL discount the possibility that variability of $\alpha$ could modify charged black hole properties enough to overturn their conclusion.

But DDL's implicit assumption that the mass of a black hole in the expanding universe is constant is incorrect.  Expansion makes the black hole environment  time dependent, and there is no reason for mass (Hamiltonian) to be conserved (even if no radiation flows in or out). The expansion timescale is generally long compared to the hole's dynamical timescale, so the process is adiabatic. And as pointed out by Fairbairn and Tytgat~\cite{FT} and Flambaum~\cite{flam}, it is rather the black hole entropy (or horizon area) which is unchanged under these circumstances (adiabatic invariance of the black hole area was established earlier by Mayo~\cite{mayo} and by one of us~\cite{beka}).   Consulting \Eq{newarea} we see that the black hole parameters should evolve with $b\sim (1-\sigma)^{G/G+\kappa^2}$.   Then \Eqs{sigma}{M} determine the dependence $M(e^{\psi_c})$, and hence $M$'s temporal variation.  Similar remarks are made by Fairbairn and Titgart on the basis of the $\kappa^2=G$ case of the solution (\ref{psi})-(\ref{e2B}).  Of course, once radiation processes are allowed for, the overall entropy must increase.  Thus the validity of the generalized second law is not endangered by growth of $\alpha$.
 
DDL also warn that systematically growing $\alpha$ will eventually bring the black hole to the point of becoming a naked singularity.  In the RN case (and in their language) this happens when the growing charge reaches $\surd G M$.  In our framework the question is whether inequality~(\ref{bound}) will fail because $e^{-\psi_c}$ decreases as $\alpha^{-1/2}$.  But can the limiting case of the inequality be reached in view of the adiabatic invariance of $\mathcal{A}$? According to \Eqs{newarea}{sigma}
\beq
\mathcal{A}\propto \frac{q^{\frac{4G}{G+\kappa^2}}\  b^{-2\frac{G-\kappa^2}{G+\kappa^2}}}{\left(\sqrt{1+q^2/8b^2}-1\right)^\frac{2G}{G+\kappa^2}}\ .
\eeq
It may be seen that as long as $q^2$ is finite, $b$ cannot approach zero while keeping $\mathcal{A}$ constant.  Now the mentioned limiting case is attained as $b\to 0$ (see discussion following \Eq{bound}). Hence, the disaster envisaged by DDL cannot take place while $\alpha$ is finite.  

The above ignores the effects of radiation.  Suppose the hole is so small (and hot)  that Hawking emission dominates both radiation accretion and $M$'s growth due to $\alpha$ evolution, but not hot enough to emit charges, so that $Q$ remains fixed.  Then it seems that  bound (\ref{bound}) will eventually be surpassed since $e^{-\psi_c}$ decreases.  Let us check if the hole can reach  the limiting point of the inequality  in a finite time.  

The energy emission rate $|\dot M|$ should, on physical grounds, be proportional to $\mathcal{A}(b) T_{BH}{}^4$. We see from \Eqs{newarea}{TBH} that near the limiting point ($b=0$)
\beq
\dot M= -\textrm{const.}\times (1-\sigma)^\frac{6G}{G+\kappa^2} b^{-2}= -\textrm{const.}\times  b^\frac{4G-\kappa^2}{G+\kappa^2},
\eeq
where the second equality follows from the fact (Sec.~\ref{sec:extreme})  that near $b=0$,  $b\sim 1-\sigma$.   Using \Eq{dbdM} we convert this into
\beq
\frac{db}{dt}=-\textrm{const.}\times  \frac{G(G+\kappa^2)}{\kappa^2}\, b^\frac{4G-2\kappa^2}{G+\kappa^2}.
\eeq
For $\kappa^2\leq G$ the integral 
\beq
\int_0^b {b'}^\frac{2\kappa^2-4G}{G+\kappa^2}\, db'
 \eeq
diverges at the lower limit, which shows that Hawking radiation cannot bring the hole to the limiting point in a finite time.  But for  $\kappa^2> G$ the integral converges.  But before concluding that it takes but a finite time for $b$ to shrink to zero and for the hole to achieve $|Q|=  \sqrt{G+\kappa^2}\,M e^{-\psi_c}$ and become a naked singularity even without help from varying $\alpha$, we should recall our assumption that no charged particles are emitted.   Sufficiently near $b=0$ the temperature diverges as $b^{(G-\kappa^2)/(G+\kappa^2)}$ so before the dangerous point the hole will begin to emit charged particles, no matter how massive they are.  The consequent decrease of $Q$ may steer the hole away from the limiting point. 

\section{Summary and conclusions}

A spacetime variable $\alpha$ modifies Maxwellian electromagnetism.  It thus modifies the structure of charged black holes in general relativity.  Here we have derived \emph{ab initio} the unique family of spherical static charged black holes in the framework of $\alpha$ variability proposed by one of us~\cite{bek}.  This family coincides with a one-parameter extension of the dilatonic black holes~\cite{GM,ghs,CKMO,FT}.  In contrast with the classic Reissner-Nordstr\"om black holes,  varying $\alpha$ charged black holes lack an inner horizon; one can take the view that variability of $\alpha$, however weak, destabilizes the inner (Cauchy) horizon to a singularity.

Our charged black hole metric has two additional sectors.  One describes the static asymptotically flat spacetime around a charged naked timelike singularity.  The last sector represents a static finite spacetime lying between two timelike singularities of zero area and bearing opposite charges.  This last configuration can occur only when the $\kappa^2$ parameter takes on one of an infinite set of rational values.

Charged black holes in varying $\alpha$ theory obey the ``no hair'' principle; they are fully determined by  the mass $M$ and conserved charge $Q$ of the black hole (and by the asymptotic value of the $\alpha$ field, which is nothing but the coeval cosmological value of $\alpha$).  The allowed range of the charge-to-mass ratio $Q/M$ is, however, somewhat different from that in the Reissner-Nordstr\"om case, and depends both on the $\kappa^2$ parameter and on the cosmological $\alpha$.  The area of the horizon of the geometry tends to zero at the largest allowed $|Q|/M$, and this limiting case of the solution is not a black hole.  Nearly extremal black holes have the property that the $\alpha$ field is a power of the square of the time Killing vector.  For the special value $\kappa^2=G/2$, an externally sourced magnetic field will be uniform in such a black hole's vicinity (as judged from infinity).  

We have here calculated anew the black hole thermodynamic functions in the face of varying $\alpha$.  In particular, we present a trick which enables an otherwise intricate calculation of the electric potential to be carried out almost trivially.  The $\alpha$ dependence of the black hole thermodynamic functions makes it tempting to suppose that black hole thermodynamics may restrict $\alpha$ variability in the expanding universe.  Such a claim was urged by Davies, Davis and Lineware~\cite{davies} because the black hole entropy for the Reissner-Nordstr\"om black hole \emph{would seem to decrease} as $\alpha$ increases. We reiterate, with Fairbairn and Tytgat~\cite{FT} and with Flambaum~\cite{flam}, that in view of adiabatic invariance of the black hole entropy, the generalized second law is not endangered by $\alpha$ growth in cosmology.  Adiabatic invariance is also sufficient to prevent a charged black hole from evolving, due to cosmological $\alpha$ growth, into the extremal state of vanishing horizon area.  However, we find that when $\kappa^2>G$, even in the absence of cosmological $\alpha$ growth, Hawking radiation of \emph{neutral} particles tends to drive a charged black hole to the vanishing horizon area state in a finite time.   However, it seems likely that late emission of charged particles due to a rising black hole temperature will prevent violation of cosmic censorship.

\acknowledgments

JDB thanks Oleg Lunin, Victor Flambaum and David Garfinkle for useful remarks.

\section*{Appendix: Magnetic black holes}
\label{sec:magnetic}
\renewcommand{\theequation}{A.\arabic{equation}}
\setcounter{equation}{0}

The duality principle informs us that there should also exist purely magnetically charged black holes in the theory of varying $\alpha$.
We again restrict attention to static spherically symmetric solutions.
In this case we expect the only nonvanishing component of the electromagnetic field tensor to be $F_{\theta\varphi}$, which corresponds to a radial magnetic field.  In terms of dual fields this is $^*F^{tr}$, so that instead of \Eq{Maxwell} we have here
\beq
(r^2 e^{A+3B}\ {}^*F^{tr})'=0.
\eeq
By analogy with \Eq{Field} we have the solution
\beq
F_{\theta\varphi}=-e^{A+3B} r^2 sin\theta\ ^*F^{tr}=P \sin\theta
\eeq
where $P$ is an integration constant, the magnetic monopole.

Forming $F^{\alpha\beta}F_{\alpha\beta}$ we now find, instead of \Eq{sc2},
\beq
(e^{A+B} r^2 \psi')'=-\kappa^2 P^2 {e^{-2\psi+A-B}\over r^2}.
\label{sc4}
\eeq
And calculating anew the electromagnetic stress-energy tensor we find the Einstein equations
\bea
tt:\qquad\ A'' +{2A'\over r}+ A'B'+A'^2 = K\qquad\qquad\qquad&&
\label{Ett'}
\\
rr:\ A''+2B''+{2B' \over r}-A'B'+A'^2= K-{2G\over \kappa^2}\psi'^2\quad&&
\label{Err'}
\\
\theta\theta:\qquad  B''+ A'B'+B'^2+{A'\over r}+{3B' \over r}=- K\quad\qquad&&
\label{Ethth'}
\\
K\equiv GP^2{e^{-2\psi-2B}\over r^4}\qquad\qquad\qquad\qquad&&
\eea

Comparing Eqs.~(\ref{sc4})-(\ref{Ethth'}) with Eqs.~(\ref{sc2})-(\ref{Ethth}) we notice that they differ only by the replacement $Q\hookrightarrow P$ and $\psi \hookrightarrow -\psi$.  Accordingly we can immediately write down the solution for the magnetic black hole:
the metric is still given by \Eqs{e2A}{e2B} with $M$ and $b$ defined by \Eqs{M}{b1} or \Eq{b} with $Q\hookrightarrow P$, while the scalar field takes the form
\beq
\psi=\psi_c-{\kappa^2\over G+\kappa^2}\ln
\left(\frac{1-\sigma}{1-\sigma\big({r-b\over r+b}  \big)^2  }\right)
\label{psi*}
\eeq
where we have turned around again the sign of $\psi_c$ since it is by definition the asymptotic value of the scalar field. By the same token, the sign of $\psi_c$ is to be retained unchanged in places like \Eq{q}, (\ref{psi1})  and (\ref{psi2}).
The discussion in Secs.~\ref{sec:dilaton}-\ref{sec:thermo} can be taken over almost verbatim: it is now $|P|$, the conserved monopole, that is restricted by \Eq{bound}, and the observable magnetic monopole is $e^{-2\psi_c} P$.

\end{document}